\documentclass[twocolumn]{aastex62}

\def\ltsima{$\; \buildrel < \over \sim\;$}
\def\ltsim{\lower.5ex\hbox{\ltsima}}
\def\gtsima{$\; \buildrel > \over\sim \;$}
\def\gtsim{\lower.5ex\hbox{\gtsima}}
\def\ms{$M_{\odot}$ }
\def\msp{$M_{\odot}$}

\received{xxx}
\revised{xxx}
\accepted{xxx}

\shorttitle{Remarkable migration of the solar system}
\shortauthors{Tsujimoto \& Baba}

\begin{document}

\title{Remarkable migration of the solar system from the innermost Galactic disk; a wander, a wobble, and a climate catastrophe on the Earth}

\correspondingauthor{Takuji Tsujimoto}
\email{taku.tsujimoto@nao.ac.jp}

\author[0000-0002-9397-3658]{Takuji Tsujimoto}

\author[0000-0002-2154-8740]{Junichi Baba}

\affiliation{National Astronomical Observatory of Japan, Mitaka, Tokyo 181-8588, Japan}



\begin{abstract}
Recent knowledge of Galactic dynamics suggests that stars radially move on the disk when they encounter transient spiral arms that are naturally generated during the process of disk formation. We argue that a large movement of the solar system from the innermost disk over its lifetime is inferred from a comparison of the solar composition with those of solar twins within the Galactic chemical evolution framework. The implied metal-rich environment at the Sun's birthplace and formation time is supported by measured silicon isotopic ratios in presolar silicon carbide grains. We perform numerical simulations of the dynamical evolution of disk stars in a Milky Way-like galaxy to identify the lifetime trajectory of the solar system. We find that a solar system born in the proximity of the Galactic bulge could travel to the current locus by the effect of radial migration induced by several major encounters with spiral arms. The frequent feature we identify is the repeated passages of stars inside the same spiral arm owing to the wobble of stars traveling in and out of the spiral arms. We predict that such episodes are evidenced in the Earth's geological history as snowball Earth and that their occurrence times are within our prediction. In particular, the stellar motion that vertically oscillates during passages through spiral arms occasionally leads to a split into two discrete passage episodes with an interval of several tens of Myr, implying two relevant snowball Earth events that occurred in rapid succession ($\sim$7.2 and 6.5 hundred Myr ago). 
\end{abstract}

\keywords{Earth atmosphere (437); Galaxy chemical evolution (580); Galaxy dynamics (591); Milky Way disk (1050); Solar abundances (1474); Spiral arms (1559); Stellar abundances (1577); Stellar motion (1615)}


\section{Introduction}

The origin and evolution of the solar system bearing our Earth from a galactic perspective are key questions in astrophysics that remain enigmatic. Meanwhile, the view of the chemodynamic evolution of the Galactic disk has been remarkably revised. This updated perspective has been facilitated by the improved understanding of Galactic dynamics guided by two distinct structures of the Milky Way: its spiral arms and the Galactic bar. Spiral arms, which are recurrent and transient, make stars migrate on the disk in the radial direction via an exchange of angular momentum \citep{Sellwood_02, Roskar_08, Grand_12a, Roskar_12, Baba_13}, while the classical long-lived rigidly rotating models have a similar dynamical effect around the location of corotation \citep{Lepine_03, Daniel_15}. Similarly, galactic bars, which are a characteristic feature among the majority of local disk galaxies, are also suggested to induce the radial movement of stars as an outcome of bar resonances \citep{Minchev_10, Minchev_11, Halle_15, Chiba_19, Khoperskov_20}. This so-called radial migration of stars predicts that the birthplaces of a considerable number of disk stars are different from their current positions. According to this prediction, the stars in the solar vicinity represent the mixture of stars born at various Galactocentric distances ($R_{\rm GC}$) over the disk. The impact of this migration on the local chemistry would be of significance if the stars born outside the solar vicinity follow chemical enrichment tracks that deviate from the locally expected tracks. In fact, the chemical evolution of the disk differs in accordance with $R_{\rm GC}$, which is observationally evidenced by current radial abundance gradients \citep[e.g.,][]{Shaver_83}. This observation can be understood under the scenario for galaxy formation in which the inner region formed faster and became more metal-rich than the outer region \citep{Chiappini_01}. This scenario, together with the idea of radial migration, nicely resolves the enigmatic problems regarding the local chemistry in the solar vicinity: the presence of super-metal-rich stars, which are unpredictable as a consequence of in situ star formation, and the widespread age-metallicity relation among stars \citep{Roskar_08, Schonrich_09}, as well as a positively skewed metallicity distribution in the outer disk \citep{Hayden_15}. 

The established new paradigm synchronizes with the orbital history of the solar system: it has been argued that the solar system migrated from the inner region of the disk in terms of its metallicity \citep{Wielen_96, Clayton_97, Nieva_12, Minchev_18, Feltzing_20}. In general, the abundances of heavy elements in interstellar matter increase with time via the cycles of star birth and death, which are recorded in the abundances of long-lived stars registered at birth. Accordingly, if the Sun was born at its current position ($R_{\rm GC}\approx$ 8 kpc), the local metallicity at the present time should be higher than the solar metallicity as an outcome of heavy element enrichment over 4.56 Gyr. However, there is little evidence for the presence of supersolar gas in the solar vicinity. Nearby young stars together with the H II region population approximately have solar metallicity \citep[e.g.,][]{Nieva_12}. A natural interpretation of these observations is that the Sun was born in the inner disk, where the metallicity was already solar approximately 4.6 Gyr ago due to fast chemical enrichment, while the local gas has only recently reached solar metallicity. It should be noted that this implied Sun's displacement is no exception since many stars in the solar vicinity are suggest to have migrated from smaller radii \citep{Hayden_15, Loebman_16}.

\section{Evaluation of the Sun's birthplace}

We can infer the Sun's birthplace ($R_{\rm birth, Sun}$) by comparing the predicted current [Fe/H] values with the observed [Fe/H]-$R_{\rm GC}$ relation (i.e., the radial [Fe/H] gradient) \citep{Genovali_14}, using the models which follow an evolutionary path passing [Fe/H] $\approx 0$ approximately 4.6 Gyr ago. Here, the models are designed for corresponding values of $R_{\rm GC}$ shorter than 8 kpc to realize more efficient chemical enrichment, leading to a faster increase in metallicity and higher metallicity compared to the model for the solar vicinity \citep{Tsujimoto_12}. According to this comparison, $R_{\rm birth, Sun}$ is found to be \ltsim 5 kpc. This value is smaller than the estimate (= 6.6 kpc) by \citet{Wielen_96}; however, the possibility of such a large movement of the Sun has been suggested \citep{Kaib_11}. Besides, the following observations will drive the Sun's birthplace farther from the current locus.

\begin{figure}[t]
    \vspace{0.2cm}
	\includegraphics[width=\columnwidth]{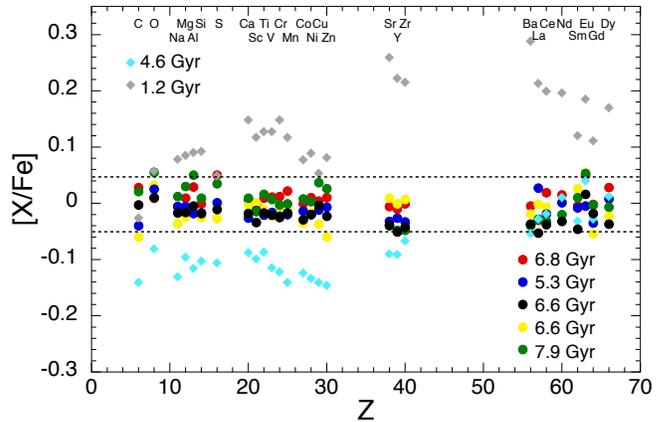}
\caption{Elemental abundance patterns of solar twin stars for the elements ranging over 6 $\leq$$Z$$\leq$ 66. From 79 solar twins \citep{Bedell_18}, we select 5 stars that show the five smallest deviations from the solar pattern (i.e., [X/Fe]=0) within an offset of $\sim$0.05 dex for all elements. Here, we remove the element Pr ($Z$=59) from this analysis since all stars exhibit a large deviation from the solar Pr/Fe ratio. The stellar ages of the five selected stars are 6.8, 5.3, 6.6, 6.6, and 7.9 Gyr old in order of the degree of smallest deviation; all of these stars are older than the Sun. Regarding the other stars, the results of two stars with a young age (1.2 Gyr) and an age similar to the Sun (4.6 Gyr) are also shown.}
\end{figure}

The detailed elemental abundance pattern of a star functions as its identify, and thus, the similarity of the patterns among stars can be associated with the commonality of their birthplace \citep{Freeman_02}. Locally, there exist many stars that are nearly identical to the Sun, the so-called solar twins, which exhibit stellar atmospheric characteristics quite similar to the solar values, i.e., an effective temperature within $\pm$100 K and a logarithmic surface gravity and [Fe/H] ratio of $\pm$0.1 dex. Thanks to such a close resemblance, the stellar ages of solar twins can be precisely determined with a typical uncertainty of 0.4 Gyr together with high-quality (an error of $<$ 0.01 dex) chemical abundances \citep{Spina_18}. From a list of these data, we first realize that the ages of solar twins are widely distributed over $0-10$ Gyr, implying that locally identified solar twins might represent the assembly of stars migrating from various $R_{\rm GC}$ within the inner disk. With our expectation regarding the similarity of abundance patterns between the Sun and its solar twins tagged with ages close to that of the Sun, we compare the chemical abundances of 28 elements from carbon ($Z$=6) to dysprosium ($Z$=66) for 79 solar twins \citep{Bedell_18}. We find that the result runs contrary to our expectation: the stars exhibiting the least deviation from the solar abundance pattern are a cluster of older stars having ages of approximately 7 Gyr (Fig.~1). Here, we stress that the chemical compositions of these older stars perfectly agree with that of the Sun within 0.05 dex for all elements. 

This revealed association of the Sun with older solar twins suggests a common birthplace, specifically the region closer than previously deduced $R_{\rm birth, Sun}$ to the Galactic center; in this common region, interstellar matter was enriched quickly to [Fe/H] $\approx$ 0 approximately 7 Gyr ago. For the subsequent chemical evolution up to 4.6 Gyr ago, it is possible to consider that [Fe/H] remained almost flat. However, a more plausible scenario might be that [Fe/H] continued to increase by $\sim$0.1$-$0.2 dex on average with a patchy metallicity distribution including spots of [Fe/H] $\approx$ 0; this case implies that the solar system formed within interstellar matter (gas + dust), which was surrounded by more metal-rich ([Fe/H] $>$ 0) stars. This theoretical scheme is strongly supported by the presence of presolar silicon carbide (SiC) grains ($\mu$m to sub-$\mu$m-sized minerals rarely found in primitive meteorites) having $^{29}$Si/$^{28}$Si and $^{30}$Si/$^{28}$Si isotopic ratios larger than those in solar material; these grains are considered to be the outcome of dust that was ejected from stars (asymptotic giant branch stars) that were more metal-rich than the Sun \citep{Clayton_97}. 

Therefore, it would be more plausible to consider that $R_{\rm birth, Sun}$ may correspond to the radial position where interstellar matter was enriched to [Fe/H]$\approx$0 approximately 7 Gyr ago, not 4.6 Gyr ago, as a mean metallicity with some variation in [Fe/H] at each time. From a purely theoretical approach, it is not possible to determine $R_{\rm birth, Sun}$ by predicting the current [Fe/H] since the continuous star formation up to the present day is unlikely under such circumstances where chemical enrichment is realized as fast as the formation of the Galactic bulge or the thick disk \citep{Tsujimoto_12}. However, we can at least constrain $R_{\rm birth, Sun}$ to be smaller than $\sim 5$ kpc. On the other hand, an analysis of the temporal evolution of the radial [Fe/H] distribution in gas deduced from observational inputs provides a radius corresponding to [Fe/H] $\approx$ 0 at 7 Gyr ago, though this estimate involves large uncertainties. According to the results from several works published to date, 3 \ltsim $R_{\rm birth, Sun}$ \ltsim 6 kpc has been inferred \citep{Minchev_18, Feltzing_20}. Ultimately, the argument based on the chemistry of the Sun predicts a displacement of the solar system as large as $\Delta R$ \gtsim 3 kpc over the Sun's lifetime. This finding should be further investigated from the perspective of Galactic dynamics.

\section{Numerical Simulations}

We next perform orbital calculations of stars in the modeled galaxy, including the structures of the spiral arms and the Galactic bar, for which we adopt the dynamically evolving spiral arm model and the slow-down bar model, respectively. Both non-axisymmetric structures drive stars to radially move through resonant interactions. 

\subsection{Models}

We first construct an axisymmetric model of a stellar disk with a bulge embedded in a dark matter (DM) halo. These structures are created by a static potential as follows. The stellar disk is composed of two separate populations, the so-called thick and thin disks. Both disks are assumed to follow an exponential profile with a mass of $4.5 \times 10^{10} \ (8.0 \times 10^9)$ \msp, a scale length of 2.6 (2.0) kpc, and a scale height of 300 (900) pc for the thin (thick) disk, and both disks are modeled by combining three separate Miyamoto-Nagai potentials \citep{Smith_15}. The density profiles of the bulge and DM halo follow the Hernquist profile \citep{Hernquist_90} and the Navarro-Frenk-White profile, respectively. Here, the mass and scale length of the bulge are set to be 4.3 $\times 10^9$ \ms and 350 pc, respectively. For the DM halo, we assume that the mass, scale length, and concentration parameter are $9.5 \times 10^{11}$ \msp, 15.62 kpc, and 13.7, respectively.

Then, we establish the non-axisymmetric structures. For spiral arms, we adopt the dynamically evolving spiral arm model, which shows the temporal changes in the pitch angle and the amplitude \citep{Hunt_18} and has a functional form of the gravitational potential \citep{Cox_02}. We assume that the amplitude increases with time for the first 0.5 Gyr and levels off afterwards with a density enhancement of 36 \% with respect to the disk. This adopted amplitude is approximately median among the observed values ($\sim$ 15$-$60\%) for external galaxies \citep{Rix_95}. For the lifetime and recurrence interval of spiral arms, we consider two cases with (200, 800) and (180, 720) in units of Myr \citep{Hunt_18}. Here, we assume a Gaussian for the temporal evolution of spiral arms. The lifetime is defined as its dispersion ($\sigma$) and after the elapse of $2\times\sigma$ (i.e., 400 or 360 Myr), when the amplitude decays to $1/e^2$ of its peak, a spiral arm is assumed to be newly born. With these adopted timescales of lifetime and interval, spiral arms periodically emerge and grow from the same specific phase every time in our simulations. Such periodicity seemingly contradicts a stochastic phenomenon of dynamic arms \citep[e.g.,][]{Roskar_12}. However, in barred galaxies like our own, it has been suggested that the emergence of spiral arms becomes quasi-periodic via an interaction with a bar \citep{Sellwood_88, Baba_15}. The number of arms ($m$) is set to 2. We assume that a scale length of the radial density profile of spiral arm is 7.0 kpc. 

An important feature of dynamic spiral arms is that their spiral pattern speeds ($\Omega_p$) are not constant but that spirals corotate with the galactic rotation: $\Omega_p \propto V_{\rm cir}/R_{\rm GC}$, where $V_{\rm cir}$ is the circular velocity \citep{Grand_12a, Baba_13, Kawata_14}. Therefore, the pattern speed decreases with radius in a manner similar to galactic rotation for any radii and thereby the resonant radii emerge throughout the entire disk, which leads to an efficient migration of stars, compared to the time-dependent spiral model with a single spiral pattern speed \citep{Martinez_17}. Such dynamic spiral nature is identified by $N$-body barred spiral galaxies \citep[][]{Grand_12b, Baba_15}. In addition, the phenomena of such corotating spiral arms have been observationally confirmed in some nearby spiral galaxies \citep[e.g.,][]{Meidt_09}. 

For the Galactic bar, we adopt the slow-down bar model. The gravitational potential of the bar is given by a functional form \citep{Binney_18}, and the amplitude is assumed to be constant. We assume that the rate of decrease in the bar pattern speed is exponential \citep{Halle_15, Khoperskov_20} with a timescale of 1 Gyr. This timescale is consistent with the values ($\sim$1$-$3 Gyr) obtained by $N$-body simulations of barred galaxies \citep[e.g.][]{Dubinski_09, Saha_13}, and we confirm that the results are essentially unchanged within its timescale range. For the current pattern speed, we consider 36 km s$^{-1}$ kpc$^{-1}$ \citep{Binney_20}, which corresponds to 1.25 times of $\Omega_{\rm sun}$. Here, $\Omega_{\rm sun}$ is defined as $V_{\rm cir} (R_{\rm sun})/R_{\rm sun}$, where $R_{\rm sun}$ is the present position of the Sun \citep[8.2 kpc:][]{Bland_16} and is deduced to be 29 km s$^{-1}$ kpc$^{-1}$ from $V_{\rm cir} (R_{\rm sun})$ = 238 km s$^{-1}$ \citep{Bland_16}. We assume that the Galactic bar already exists at the time of solar system formation \citep{Bovy_19}. Considering its slowing-down timescale, the pattern speed of the bar is set to 55 km s$^{-1}$ kpc$^{-1}$ at the time of solar system formation. The slow-down bar can be expected to efficiently move stars toward the outer disk through the dynamical process in which the migrating star are captured by the outward resonant radius \citep{Halle_15, Chiba_19, Khoperskov_20}, which is initially located around $R_{\rm GC}$=4 kpc outside the bar in our model. 

For the initial setup, we distribute 100,000 star particles in the distance range of 3 \ltsim $R_{\rm GC}$ \ltsim 5 kpc on the disk in the modeled galaxy. We assume that the velocity dispersion of interstellar matter at the formation of the solar system (i.e., 4.6 Gyr ago) is ($\sigma_R$, $\sigma_\psi$, $\sigma_Z$) = (12.5, 8.8, 12.5) km s$^{-1}$ according to the functional formula given by \citet{Aumer_17}. Then, adopting an asymmetric drift value of 1.9 km s$^{-1}$, which is deduced from the obtained velocity dispersion, the mean velocity is given as ($\langle V_R\rangle$, $\langle V_\psi\rangle$, $\langle V_Z\rangle$) = (0, $V_{\rm cir} (R_{\rm GC})-1.9$, 0) km s$^{-1}$, where $V_{\rm cir} (R_{\rm sun})$ is the circular velocity at radius $R_{\rm GC}$. Using these estimates, we assign the velocities to each particle to follow a Gaussian distribution. We trace the dynamical evolution of these 100,000 stars over a period of 4.6 Gyr and extract the stars, the eventual motions of which are analogous to the solar motion. For this purpose, we calculate their angular momentum ($L_Z$) together with their eccentricities ($e$) from the stellar velocities with an assumption of a flat rotation curve \citep{Arifyanto_06}. Then, we select the stars that satisfy $L_Z$ within 2033$\pm$102 kpc km s$^{-1}$ (i.e., $L_Z$ with its deviation from the solar value within 10 \%) and $e<0.1$. Here, the solar $L_Z$ ($L_{\rm Z, sun}$) is deduced from $R_{\rm sun}^2\Omega_{\rm sun}$ where $\Omega_{\rm sun}$ is taken from the observed value of 30.24 km s$^{-1}$ kpc$^{-1}$ \citep{Reid_04}. Finally, we add another condition that the vertical motion of stars with respect to the disk is approximately 100 pc. The guiding radius ($R_g$:~approximately an orbital radius) of the Sun is defined as $L_{\rm Z, sun}/V_{\rm cir} (R_{\rm sun})$, which is estimated to be 8.11$-$8.97 kpc, allowing a 10 \% uncertainty. Here, the guiding radius of the Sun (8.5$\pm$0.4 kpc), which is larger than the current distance of 8.2 kpc, is attributed to the fact that the Sun currently approaches the pericenter of the Galaxy.

\subsection{Results}

\begin{figure*}[t]
    \vspace{0.2cm}
    \hspace{0.5cm}
	\includegraphics[angle=90, width=2.0\columnwidth]{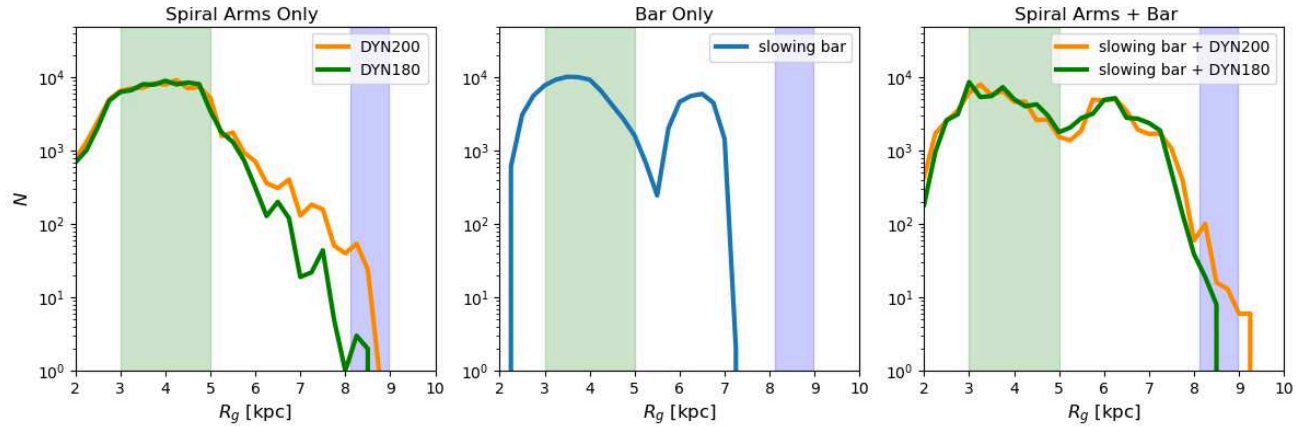}
\caption{Eventual $R_g$ distribution for the stars initially set at Galactocentric distances within 3\ltsim$R_{\rm GC}$\ltsim5 kpc as a result of radial migration over 4.6 Gyr triggered by spiral arms, the Galactic bar, and both. $R_g$ is the stellar guiding radii estimated from stellar angular momentum and circular velocity, and is approximately equivalent to $R_{\rm GC}$. The range of $R_{\rm birth, Sun}$ and the guiding radius of the Sun at present are indicated by the green and purple shaded zones, respectively. ({\it left panel}) The results with dynamic spiral arms (DYN) only. ({\it middle panel}) The result with the slow-down bar only. ({\it right panel}) The results with the combination of dynamic spiral arms and a slow-down bar.}
\end{figure*}

We show the final distance distribution for these stars in Figure 2. We find that dynamic arms alone can transport some of stars $R_{\rm GC}$ \gtsim 8 kpc, while the bar alone can hardly transport stars beyond $R_{\rm GC}$ $\approx$ 7 kpc. Here, we note that  a small fraction of stars in the range of 3 \ltsim $R_{\rm GC}$ \ltsim 5 kpc belong to the bar end and thus are incapable of migrating outward. In tandem with dynamic arms, the slow-down bar enhances the fraction of stars that eventually migrate up to 6 \ltsim $R_{\rm GC}$ \ltsim 8 kpc; however, its effect is found to be small for the migration over distances of $R_{\rm GC}$ \gtsim 8 kpc. Accordingly, the sporadic resonant interactions of the solar system with dynamic arms are essential to its large movement (such as \gtsim3 kpc) within 4.6 Gyr \citep{Kaib_11}.

For stars that eventually have a guiding radius of $\sim$8.5 kpc and thus could be candidates reflecting the orbital history of the solar system, we trace their trajectories and examine how each star's $R_{\rm GC}$ changes with time in association with the history of encounters with spiral arms. Here, we identify the arm passage times as the density excess peaks by the spiral arms at the location of a moving star. Two representative cases together with an averaged variation are presented in Figure 3. These results demonstrate that stars take an asymptotic pathway to the outer disk while wandering inward/outward on the disk with a small $R_{\rm GC}$ oscillation owing to an epicycle motion. In the meantime, stars typically experience several encounters with spiral arms over 4.6 Gyr. 

A closer look at these arm-passage events clarifies that some of these events cause stars to reside inside the arms for a relatively long time; this long residence time is identified as a broad spiral arm density peak. This feature results from the wobble of stars traveling in and out of spiral arms due to a small relative rotational velocity between the passing stars and dynamic arms. If the solar system experienced such long passages, we could expect that the Earth suffered enormous irradiation, including cosmic rays from supernovae, during this residence period since spiral arms are the cradles of massive baby stars. The fossil record of this crisis must be, in some cases, catastrophic climate change on Earth \citep{Shaviv_02, Gies_05}. Historical events during which all liquid water on the surface of the Earth was frozen, the so-called snowball Earth \citep{Kirschvink_92, Hoffman_98}, have occurred at least three times, i.e., 2.43, 0.717, and 0.650 Gyr ago \citep{Hoffman_19}. Although the trigger responsible for such an event is still under debate, including a hypothesis claiming a connection of the first event (the Huronian glaciation) to the Great Oxidation Event \citep{Kirschvink_00}, we propose that the mechanism for snowball Earth could be wobbly passages of the solar system through spiral arms based on the hypothetical strong cooling effect owing to an increase in cloudiness seeded by an enormous influx of cosmic rays \citep{Svensmark_07}. 

In particular, the unique and puzzling appearance of two snowball Earth events (the Sturtian and Marinoan glaciations) in rapid succession can be explained within our framework, suggesting a vertically oscillating motion of stars passing through arms, which occasionally results in two separate spiral arm passage episodes with an interval of several tens of Myr. Note that snowball Earth events might not be expected for the early Earth because of the chaotic bombardment phase (the lookback time $t_{\rm L}$ \gtsim 3.9 Gyr) or the blocking of cosmic rays by high solar magnetic activity ($t_{\rm L}$ \gtsim 2.6 Gyr). Accordingly, our model shows broad coincidence between the times of spiral arm passages and snowball Earth events for $t_{\rm L}$ \ltsim 2.6 Gyr. This suggests that the orbital history of the solar system in association with the Earth's geological history could be the sole witness to the properties of dynamically evolving spiral arms that has enabled a remarkable migration.

\section{Summary}

\begin{figure*}[t]
	\hspace{0.5cm}
    \vspace{0.2cm}
	\includegraphics[width=1.9\columnwidth]{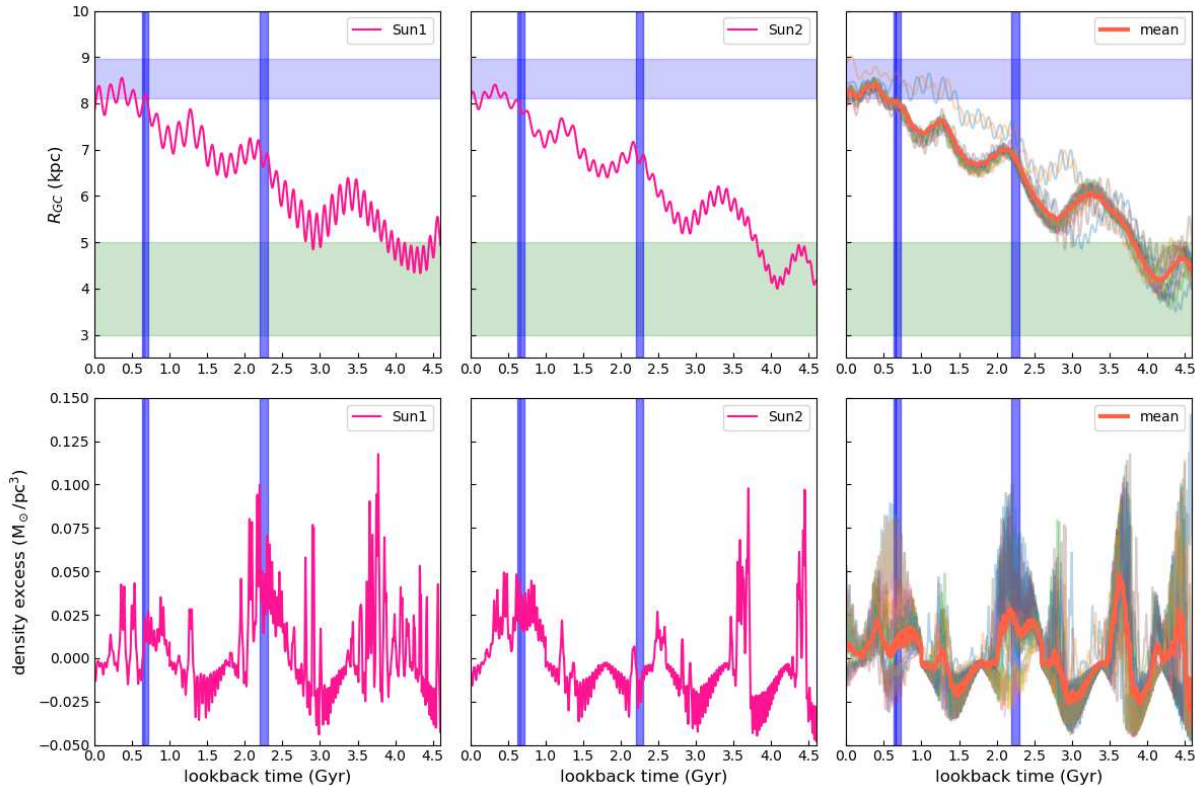}
	\vspace{0.0cm}
\caption{Trajectory and spiral arm passage history of the stars whose dynamical histories can be analogous to that of the solar system. ({\it upper panels}) The change in $R_{\rm GC}$ as a function of the lookback time for the stars selected from the model with DYN (a lifetime=200 Myr) in the right panel of Figure 2. These results demonstrate two representative cases (left, middle) and the average variation in $R_{\rm GC}$ for 22 stars that reach the Sun's orbital zone (orange line) together with individual ones denoted by various colored lines (right). ({\it lower panels}) History of the spiral arm density excess at the time-varying locations of migrating stars for two individual cases corresponding to the upper panels (left, middle), and the averaged result with individual variations (right). Each peak can be associated with an encounter event between a star and a spiral arm. 
The recorded times of snowball Earth \citep{Hoffman_19} are indicated by the blue zones: the Huronian glaciation (2.43 Gyr ago) and the Sturtian and Marinoan glaciations (0.717 and 0.650 Gyr ago).}
\end{figure*}

We argue that the solar system was born far away from its current location based on the evidence imprinted in detailed chemical compositions of solar twins which are tied up with the precise estimate of their ages. This finding is advanced to investigations by numerical simulations that reveal the dynamical history of the solar system over its lifetime that is essentially promoted by the interactions with the dynamically evolving spiral arms, as previously claimed. A close scrutiny of stellar motion leads to our prediction that the episodes of spiral arm passages could induce catastrophic climate change on Earth as an outcome of the long residence inside the arms of the solar system, which results from the characteristic motion of passing stars that wobble and travel in and out of the arms. 

Such a long travel along the arms could be attributed to the effect of winding of dynamic arms which make corotation appear over a wide radial range within arms \citep{Grand_12b, Baba_13}. On the other hand, the rigidly rotating spiral models predict a shorter stay of stars in regions of high density in the arms at their passages. Since a likelihood that the Earth experience the catastrophic climate change could be associated with the passing period along the arms, a weak connection between spiral arm passages and snowball Earths is expected for the the rigidly rotating spiral models. However, dynamic spiral arms still need to be investigated by the models closer to the realistic spiral arms in a quasi-periodic state including a stochastic mode, updated from our models that simplify the complex feature of spiral arms to explore the dynamic arm-snowball Earth connection in a more convincing way.

\vspace{-0.5cm}
\acknowledgements

The authors thank our anonymous referee for the useful comments and suggestions. T.T. thanks Kunio Inoue for discussions which inspired the idea of this work. This work was supported by JSPS KAKENHI Grant Numbers 17H02870, 18K03711, 18H01258, 18H01248, 19H05811, and 19KK0080.

\end{document}